\documentstyle[aps,prl,twocolumn,floats]{revtex}

\date{August 16, 2000}
\tighten

\begin{document}
\draft
\newcommand\lsim{\mathrel{\rlap{\lower4pt\hbox{\hskip1pt$\sim$}}
    \raise1pt\hbox{$<$}}}
\newcommand\gsim{\mathrel{\rlap{\lower4pt\hbox{\hskip1pt$\sim$}}
    \raise1pt\hbox{$>$}}}

%%%%%%%%%%%%%%%%%%%%%%%%%%%%%%%%%%%%%%%%%%%%%%%%%%%%%%%%%%%%%%%%%%%%%%%%%%

\title{Cosmology with Varying Constants}

\author{C. J. A. P. Martins\thanks{Electronic address:
C.J.A.P.Martins\,@\,damtp.cam.ac.uk}}

\address{Department of Applied Mathematics and Theoretical Physics\\
Centre for Mathematical Sciences, University of Cambridge\\
Wilberforce Road, Cambridge CB3 0WA, U.K.}

\address{and Centro de Astrof\'{\i}sica, Universidade do Porto\\
Rua das Estrelas s/n, 4150-762 Porto, Portugal}

\maketitle
\begin{abstract}

{I motivate and discuss some recent work on theories with varying
constants, and consider some possible observational consequences and tests.
Particular emphasis is given to models which can (almost) exactly mimic
the predictions of standard inflationary models.}

\end{abstract} 

%%%%%%%%%%%%%%%%%%%%%%%%%%%%%%%%%%%%%%%%%%%%%%%%%%%%%%%%%%%%%%%%%%%%%%%%%%
\section{Motivation}
\label{intro} 

According to our present understanding, higher-dimensional
theories \cite{polc}, are thought to be required to provide a consistent
unification of the know fundamental interactions of nature. Even though
there is at present no robust ideas about how one can go from these theories
to our familiar low-energy $(3+1)$ spacetime, it is clear that such a process
should necessarily involve two crucial mechanisms, namely {\it dimensional
reduction} and {\it compactification}.

From our present purposes, the most important consequence of this
process is that in such theories
the {\it effective} three-dimensional constants will typically be
related to the {\it true} higher-dimensional constants via the radii of the
(compact) extra dimensions \cite{banks}. Furthermore, it is well known
that these radii often have a non-trivial evolution\footnote{Indeed, this is
such a pressing question from the string theory point of view
that it has been promoted to the category of a
`problem'---the radius stabilisation problem.}. Hence
one is naturally led to the expectation of time (or even space) variations
of the `effective' coupling constants we can measure \cite{chodos,wu,kiritsis}.
This provides more than enough motivation to consider the cosmological
consequences of these variations.

%%%%%%%%%%%%%%%%%%%%%%%%%%%%%%%%%%%%%%%%%%%%%%%%%%%%%%%%%%%%%%%%%%%%%%%%%%
\section{Modelling and measurements}
\label{modelling} 

In order to do this, one must first build `toy-models' for the evolution of
the effective constants\footnote{This might at first seem a little ad-hoc,
until one realizes that there is no well-motivated particle physics model
for inflation. Hence so far so good...}.

These issues have been discussed at least since the time of Dirac \cite{dirac},
who first considered variations of the gravitational constant $G$. Almost
half a century later, Beckenstein \cite{beken} introduced
a theory with a varying
electric charge $e$.Finally, much more recently, there has been an
extraordinary growth in the interest in theories with a
varying speed of light $c$ \cite{mof,abm,am,bim,pol}. 

Before proceeding, one should recall that one can only measure dimensionless
combinations of dimensional constants \cite{abm,am}, and that such measurements
are necessarily local. For example, the statement that ``the speed of light
here is the same  as the speed of light in Andromeda'' is either a
{\it definition} or it's completely meaningless.
The paper by Albrecht \& Magueijo  in \cite{abm}
provides some further instructive examples of this point.

An important consequence of the above point is that when considering
observational tests one should focus on dimensionless quantities. The
most relevant example is that of the fine-structure constant,
\begin{equation}
\alpha\equiv\frac{e^2}{\hbar c}\, .
\label{alf}
\end{equation}
One consequence of what was said above is the fact that any evidence
for a variation in a dimensionfull quantity will necessarily be dependent on
the choice of units in which we choose to measure it. In other words, given a
theory with a varying constant (say $c$), one can always, by a suitable
re-definition of units of measurement, transform it into another theory
where another constant (say $e$) varies, and furthermore any two such
theories will be observationally indistinguishable. Hence, deciding which
system of units one adopts is, in some sense, a matter of convenience and
mathematical simplicity. It seems to be the case that theories with varying
speed of light $c$ are generally easier to work with than those with a
varying electric charge or Planck constant, although one could think
of counter-examples of this statement.

%%%%%%%%%%%%%%%%%%%%%%%%%%%%%%%%%%%%%%%%%%%%%%%%%%%%%%%%%%%%%%%%%%%%%%%%%%
\section{Observational status}
\label{observ} 

Tests for possible variations of the fine-structure
constant $\alpha$ have been carried out for a number of years, and the
current observational status is rather exciting, but also a little
bit confusing---see \cite{varsh} for a brief summary.

The best limit from
laboratory experiments (using atomic clocks) is \cite{prestage}
\begin{equation}
|{\dot \alpha}/\alpha|<3.7\times 10^{-14} yr^{-1}\, .
\label{labbound}
\end{equation}
Measurements of isotope ratios in the Oklo natural reactor provide the
strongest geophysical constraints \cite{damour},
\begin{equation}
|{\dot \alpha}/\alpha|<0.7\times 10^{-16} yr^{-1}\, ,
\label{oklobound}
\end{equation}
although
there are suggestions \cite{sisterna} that due to a number of nuclear physics
uncertainties and model dependencies a more realistic bound is
$|{\dot \alpha}/\alpha|<5\times 10^{-15} yr^{-1}$. Note that these
measurements effectively probe timescales corresponding to a cosmological
redshift of about $z\sim0.1$ (compare with astrophysical measurements below).

Three kinds of astrophysical tests have been used. Firstly, big bang
nucleosynthesis \cite{bbn}
can in principle provide rather strong constraints at very high redshifts, but
it has a strong drawback in that one is always forced to make
an assumption on how the neutron to proton mass difference depends on $\alpha$.
This is needed to estimate the effect of a varying $\alpha$ on the ${}^4He$
abundance. The abundances of the other light elements depend much less
strongly on this assumption, but on the other hand these abundances are
much less well known observationally. Hence one can only find the relatively
weak bound 
\begin{equation}
|{\Delta \alpha}/\alpha|<2\times 10^{-2}\, ,\qquad z\sim10^9-10^{10}.
\label{nuclbound}
\end{equation}

Secondly, observations of
the fine splitting of quasar doublet absorption lines
probe smaller redshifts, but should be much
more reliable. Unfortunately, the two groups which have been actively
studying this topic report different results. Webb and
collaborators \cite{webb} were the first to report a positive
result,
\begin{equation}
\Delta\alpha/\alpha=(-1.9\pm0.5)\times 10^{-5}\,,\qquad z\sim1.0-1.6
\label{webbpub}
\end{equation}
Note that
this means that $\alpha$ was {\em smaller} in the past. Recently the same
group reports two more (as yet unpublished) positive results \cite{webbnew},
$\Delta\alpha/\alpha=(-0.75\pm0.23)\times 10^{-5}$ for
redshifts $z\sim0.6-1.6$ and
$\Delta\alpha/\alpha=(-0.74\pm0.28)\times 10^{-5}$ for
redshifts $z\sim1.6-2.6$. On the other hand, Varshalovich and
collaborators \cite{varsh} report only a null result,
\begin{equation}
\Delta\alpha/\alpha=(-4.6\pm4.3\pm1.4)\times 10^{-5}\,,
\qquad z\sim2-4\,;
\label{varshbound1}
\end{equation}
the first error bar corresponds to the statistical error while the second
is the systematic one. This
corresponds to the bound
\begin{equation}
|{\dot \alpha}/\alpha|<1.4\times 10^{-14} yr^{-1}
\label{varshbound2}
\end{equation}
over a timescale of
about $10^{10}$ years. It should be emphasised that the observational
techniques used by both groups have significant differences, and it is
presently not clear how the two compare when it comes to eliminating
possible sources of systematic error. Clearly this is an issue which can only
be resolved with more and better data.

Finally, a third option is the cosmic microwave background
(CMB) \cite{steen,us}.
This probes intermediate redshifts, but has the significant advantage that
one has (or will soon have) highly accurate data.
A varying fine-structure constant changes the Thomson scattering cross
section for all interacting species, and also changes the recombination
history of Hydrogen (via changes in the energy levels and binding
energies of all species).
The authors of \cite{steen} provide an analysis of these effects and
conclude that future CMB experiments should be able to provide
constraints on a varying $\alpha$ at the recombination epoch (that is, at
redshifts $z\sim1000$) at the level of
\begin{equation}
|{\dot \alpha}/\alpha|<7\times 10^{-13} yr^{-1}\, ,
\label{cmbbound1}
\end{equation}
or equivalently
\begin{equation}
|\alpha^{-1}d\alpha/dz|<9\times 10^{-5}\, ,
\label{cmbbound1a}
\end{equation}
which seems to indicate that these constraints can only become competitive
in the near future. For example, a recent analysis \cite{us} which uses the
BOOMERanG \cite{boold} determination of the position of the first Doppler
peak find that this still allows for a variation of up to $4\%$ in the
speed of light after recombination. More recently \cite{ourfit}, it has been
shown that the BOOMERanG and MAXIMA data slightly prefer a fine-structure
constant that was smaller in the past, in agreement with quasar data.

We thus see that constraints at recent times are fairly strong, and any
drastic recent departures from the standard scenario are excluded.
On the other hand, there are no significant constraints in the
pre-nucleosynthesis era, which leaves a rather large open space for theorists
to build models---in fact, too large a space, as we'll see next.

%%%%%%%%%%%%%%%%%%%%%%%%%%%%%%%%%%%%%%%%%%%%%%%%%%%%%%%%%%%%%%%%%%%%%%%%%%
\section{Generalising general relativity}
\label{laws} 

Theories with varying constants require a generalisation of Einstein's theory
of relativity. However, there is no unique (or even preferred) way of doing
this. Hence one will be faced with the task of choosing a set of postulates
to characterise the chosen theory. An associated task is the choice of
`fundamental units' in which measurements are to be made in the theory---in
other words, the `rulers and clocks' for the theory.

In particular, one can break a number of invariance principles and
conservation laws. Examples of these include covariance and Lorentz
invariance, mass, particle number, energy and momentum conservation,
charge conservation and various energy conditions. In some ways, this is
perhaps a too drastic step (as argued in \cite{bass}, for example). On the
other hand, it does have the rather desirable feature \cite{abm}
of allowing rather simple solutions to some outstanding cosmological enigmas,
such as the horizon, flatness and cosmological constant problems.

But are these drastic steps really necessary, and what is physically the role
of a varying speed of light (or other constants)? In \cite{am} it was
conjectured that {\it any theory that reduces to General Relativity in some
appropriate limit and solves the horizon and flatness problems must necessarily
violate at least one of the following\footnote{{\rm Stronger requirements will
be needed if one wants to solve the cosmological constant problem as
well---which is something that inflation can not do.}}:
the strong energy condition, Lorentz
invariance or covariance}. Note that inflation is of course an example
of a theory which solves the horizon and flatness problems through a
violation of the strong energy condition. The above statement is a conjecture
in the sense that no rigorous `mathematical' proof was provided, although
physical arguments were discussed in \cite{am} which make it (we think)
rather plausible\footnote{On the other hand, it should be said that a number
of other people, most notably J. Magueijo, came quite close to providing
counter-examples.}.

What one can prove is, in some sense, the opposite result: No theory that
reduces to General Relativity and obeys the above three principles will be
able to solve the horizon and flatness problems, no matter what varying
constants are included in it. Physically this is because for any theory that
is subject to these four requirements one can always find a particular choice
of fundamental units that will transform it into the `standard' theory, where
there are no varying constants and the standard cosmological problems can not
be solved.

Nevertheless, if one considers that breaking Lorentz invariance in this way
is too drastic, there are alternatives. Most notably,
Bassett {\it et al.} \cite{bass} have extensively discussed the consequences of
soft Lorentz invariance breaking\footnote{This is in some sense analogous
to the concept of spontaneous symmetry breaking in particle physics. In this
analogy, the Lorentz invariance breaking process described previously would
correspond to explicit symmetry breaking.}. In particular, this will still
solve the horizon problem, but not the flatness problem (at least by itself).
As discussed in detail in \cite{bass}, there are numerous examples of
this kind of `effective' theories in other areas of physics.

Physically, theories with soft Lorentz invariance breaking will have
two (or more) natural speed parameters, and hence to the realm of two-metric
theories. For example, one can have a characteristic speed for photons and
another one for the rest of particles, or one for gravitons and another one for
the rest, or one for bosons and another one for fermions, or even (as will
be the case in the example that we will consider in some detail in the next
section) one characteristic speed for the Higgs sector and another one for
all the standard model (including gravity) particles.

%%%%%%%%%%%%%%%%%%%%%%%%%%%%%%%%%%%%%%%%%%%%%%%%%%%%%%%%%%%%%%%%%%%%%%%%%%
\section{Primordial adiabatic fluctuations from defects}
\label{adia} 

There are currently two basic classes of models that could be responsible
for producing ``seeds'' for structure formation. In the first \cite{vsh},
it is assumed that the universe
was smooth at the start of its standard evolution, and defects
were produced at one or more symmetry-breaking phase transitions which
then continuously seeded structures on a specific set of comoving
scales. In the second \cite{linde}, an inflationary
epoch is assumed to have happened before the standard evolution of the
universe began, and the corresponding primordial fluctuations were
laid out at this earlier epoch.

The main difference between these two scenarios is related to causality.
In the first case, the initial conditions for the defect network that will
be responsible for the primordial seeds are set up on a Cauchy surface
that is part of the standard history of the universe. Hence, there will
not be
any correlations between quantities defined at any two spacetime points
whose backward light cones do not intersect on that surface. On the other
hand, inflation effectively pushes this surface to much earlier times,
and if the inflationary epoch lasts long enough to solve the
well-known set of ``cosmological enigmas'' then there will be
essentially no causality constraints.

This can be seen in an alternative way by
noting that inflation can be physically
defined as an epoch when the comoving Hubble
length decreases. Hence this length starts out very large, and perturbations
can therefore be generated causally. Then inflation forces this length to
decrease enough so that, even though it grows again after inflation ends,
it's never as large (by today, say)
as the pre-inflationary era value. Note that once
the primordial fluctuations are produced they can simply freeze in comoving
coordinates and let the Hubble length shrink and then (for small enough
scales) grow past them later.

As a first step towards identifying the specific model that operated in
the early universe, one would like to be able to determine which of the two
mechanisms above (if any) was involved. Features like 
super-horizon perturbations or the so-called `Doppler peaks' \cite{gun}
on small angular scales in the CMB angular power spectrum, however, do
not provide good discriminators (at least on their
own) \cite{turok,hor,also,desi,liddle,alex}.
Gaussianity tests \cite{gauss1,gauss2,ASWA2} are in principle a better
discriminator, though it is possible to build inflationary models
that produce some forms of non-Gaussianity. Notably, it is easy to obtain
non-Gaussianity with a chi-squared distribution---an example are the so
called iso-curvature inflation models \cite{isoc}. On the other hand, if one
found non-Gaussianity in the form of line discontinuities, then it is hard
to see how cosmic strings could fail to be involved.

The above discussion shows that even though defect and inflationary models
have of course a number of distinguishing characteristics, there is a
greater overlap between them than most people would care to admit.
Moreover, it is also quite easy to obtain models where both defects and
inflation generate density fluctuations \cite{acm2,cb}.
We discuss an explicit example \cite{bim} of a model
where the primordial fluctuations are generated by a network of cosmic
defects, but are nevertheless very similar to a standard inflationary model;
the only difference between these models and the standard inflationary
scenario will be a small non-Gaussian component.

We consider a theory that contains two different speed parameters,
say $c_\phi$ 
and $c$; the first is relevant for the dynamics of the scalar field which
will produce topological defects, while 
the second is the ordinary speed of light that is relevant for gravity and
all standard model interactions.

The basic idea should now be clear. We assume that $c_\phi \gg c$ so
that the correlation length of the  network of topological defects will be
much greater than the horizon size (which is of course defined with respect
to $c$). We could, in analogy with \cite{bass}, explicitly define our
effective theory by means of an action, and postulate a relation
between the `standard' metric and the one describing the propagation of
our scalar field. However, this is not needed for the basic point we're
discussing here. Also, we concentrate on the case of
cosmic strings, whose dynamics and evolution are much better known than
those of other defects \cite{vsh,ms1,ms2,thesis}
although much of what we will discuss will apply to other defects as well.

Note that $c_\phi$ could either be a constant, with say
\begin{equation}
[g_\phi]_{00}=\, \frac{c_\phi^2}{c^2}\, g_{00}\, , 
\label{howc}
\end{equation}
or, as discussed in \cite{bass} one could set up a model such that
the two speeds are equal at very early and at recent times,
and between these two epochs there
is a period, limited by two phase transitions, where $c_\phi \gg c$. As will
become clear below, the basic mechanism will work in both cases, although
the observational constraints on it will of course be different for each
specific realization.

The evolution of the string network will be qualitatively
analogous to the standard
case \cite{vsh,ms1,ms2,thesis}, and in particular
a ``scaling'' solution will be reached
after a relatively short transient period.
Thus the long-string characteristic length scale (or ``correlation length'')
$L$ will evolve as
\begin{equation}
L=\gamma c_\phi t\, , 
\label{scall}
\end{equation}
with $\gamma={\cal O}(1)$, while the string RMS velocity will obey
\begin{equation}
v_\phi=\beta c_\phi\, , 
\label{scalv}
\end{equation}
with $\beta<1$.

Note, however, that there are a small number of important differences
relative to the standard scenario.
The first one is the most obvious: if $c_\phi \gg c$, then the string
network will be outside the horizon, measured in the usual way. Hence
these defects will induce fluctuations when they are
well outside the horizon, thus avoiding causality constraints.

On the other hand, we also expect the effect of gravitational back-reaction
to be much stronger than in the standard case \cite{model,bass}.
The general effect of the back-reaction is to reduce the
scaling density and velocity of the network relative to the standard value,
as has been discussed elsewhere \cite{model}. Thus we should expect fewer 
defects per ``$c_\phi$-horizon'', than in the standard GUT-scale case.
Having said that, it is also important to note that despite this strong
back-reaction, strings will still move relativistically. Indeed, it can be
shown \cite{model} that although back-reaction can slow strings down by
a measurable amount, only friction forces \cite{ms2,thesis} can
force the network into a strong non-relativistic regime.
Thus we expect $v_\phi$ to be somewhat lower than $c_\phi$, but still larger
than $c$.

Only in the case of monopoles, which are point-like,
one would expect the defect velocities
to drop below $c$ due to graviton radiation \cite{bass}.
This does not happen for extended objects, since their tension
naturally tends to make the dynamics take place with a characteristic
speed $c_\phi$ \cite{carter}.
This point is actually crucial, since if the network was completely
frozen while it was outside the horizon (as it happens in more standard
scenarios \cite{acm2})
then no significant perturbations would be generated.

A third important aspect, to which we shall return below,
is that the scale of the symmetry breaking, say $\Sigma$,
which produces the defects can be significantly lower than the GUT scale,
since density perturbations can grow for a longer time than usual. Indeed,
the earlier the defects are formed, the lighter they could be. Proper
normalisation of the model will produce a further constraint on $\Sigma$.

Finally, we also point out that in the scenario we have outlined above
where $c_\phi$ is a time-varying quantity which only departs from $c$
for a limited period (which is started and ended by two phase transitions),
the defects will become frozen and start to fall inside the horizon
after the second phase transition. In this case what we require is that
the defects are sufficiently outside the horizon and are relativistic
when density fluctuations in the observable scales are generated.
This will introduce additional constraints on the parameters of the model,
and in particular on the epochs at which the phase transitions take place.

The evolution of the primordial fluctuations in our model is detailed
in \cite{bim}. One obtains
a model with primordial, adiabatic ($\delta_r=4 \delta_m/3$), 
nearly Gaussian fluctuations whose primordial spectrum is of the 
Harrison-Zel'dovich form. This model is almost indistinguishable from 
the simplest inflationary models (as far as structure formation is concerned) 
except for the small non-Gaussian component which could be detected with 
future CMB experiments. The $C_l$ spectrum and the polarisation curves 
of the CMBR predicted by this model should also be identical to the ones 
predicted in the simplest inflationary models as the perturbations in the 
CMB are not generated directly by the defects.

Note that the key ingredient consists of having the
speed characterising the defect-producing scalar field much larger
than the speed characterising gravity and all standard model
particles. This provides a `violation of causality', as required
in the criterion provided by Liddle \cite{hor}.
The only distinguishing characteristic of this model, by comparison with
the simplest inflationary models, will be a small non-Gaussian signal
which could be detected by future experiments.

In fact, one can even think \cite{gauu} of an alternative model arising in
the same context, but where the defect-induced primordial fluctuations are
also Gaussian. In this case we require that the characteristic speed of the
decay products of
the defect network is much larger than the speed characterising gravity
and all the standard model particles. Thus this model
will exactly reproduce the CMB and large-scale structure
predictions of the standard inflationary models,
and the only way to identify it would be through the decay products of
the defect network involved.

We emphasise again that in open or hybrid models
of inflation defects can also be stretched outside the horizon \cite{acm2,cb}
but in this case they are frozen in
comoving coordinates so that the perturbations they induce while being
outside the horizon are negligible. 

Admittedly, these models might admittedly
seem somewhat ``unnatural'' in the context of our present
theoretical prejudices, though they are certainly not the only ones to fit
in this category \cite{turok,desi}. However, if one keeps in mind
that any fully consistent cosmological structure formation model candidate
should eventually be derivable from fundamental physics, one could argue that
at this stage they are, {\em caeteris paribus}, on the same footing as
inflation. Certainly no single fully consistent realization of an
inflationary model is known at present.

Be that as it may, however,
the fact that these examples can be constructed (and one wonders how many
more are possible) highlights the fact that extracting robust predictions
from cosmological observations is a much more difficult and subtle
task than many experimentalists (and theorists) believe.

\acknowledgements

The work presented here was done in collaboration with Pedro Avelino, and
is part of an ongoing collaboration which also involves
Graca Rocha and Pedro Viana. I thank Bruce Bassett,
Jo\~ao Magueijo, Anupam Mazumdar
and Paul Shellard for useful discussions and comments.

This work has been supported by FCT (Portugal) under
`Programa PRAXIS XXI', grant no. PRAXIS XXI/BPD/11769/97, and by
FCT/ESO, undar project ESO/PRO/1258/98.


\begin{references}
\bibitem{polc}
J. Polchinski, {\em String Theory}, Cambridge University Press (1998).
\bibitem{banks}
T. Banks, hep-th/9911067 (1999).
\bibitem{chodos}
A. Chodos and S. Detweiler, Phys. Rev. {\bf D21}, 2167 (1980);

W.J. Marciano, Phys. Rev. Lett. {\bf 52}, 489 (1984).
\bibitem{wu}
Y.S Wu and Z.W. Wang, Phys. Rev. Lett. {\bf 57}, 1978 (1986).
\bibitem{kiritsis}
E. Kiritsis, J.H.E.P. {\bf 10}, 10 (1999);

S.H.S. Alexander, hep-th/9912037 (1999).
\bibitem{dirac}
P.A.M. Dirac, Nature 139, 323 (1932).
\bibitem{beken}
J.D. Bekenstein, Phys. Rev. {\bf D25}, 1527 (1982).
\bibitem{mof}
J.W. Moffat, Int. J. Mod. Phys. {\bf D2}, 351 (1992);

J.W. Moffat, astro-ph/9811390 (1998).
\bibitem{abm}
A. Albrecht and J. Magueijo, Phys. Rev. {\bf D59}, 043516 (1999);

J.D. Barrow, Phys. Rev. {\bf D59}, 043515 (1999);

J.D. Barrow and J. Magueijo, Phys. Lett. {\bf B443}, 104 (1998).
\bibitem{am}
P.P. Avelino and C.J.A.P. Martins, Phys. Lett. {\bf B459}, 468 (1999).
\bibitem{bim}
P.P. Avelino and C.J.A.P. Martins, Phys. Rev. Lett. {\bf 85}, 1370 (2000).
\bibitem{pol}
M. Biesiada \& M. Szydlowski, Phys. Rev. {\bf D62}, 043514 (2000).
\bibitem{varsh}
D.A. Varshalovich, A.Y. Potekhin and A.V. Ivanchik, physics/0004062 (2000).
\bibitem{prestage}
J.D. Prestage, R.L. Tjoelker and L. Maleki,
Phys. Rev. Lett. {\bf 74}, 3511 (1995).
\bibitem{damour}
T. Damour and F. Dyson, Nucl. Phys. {\bf B480}, 37 (1996).
\bibitem{sisterna}
P.D. Sisterna and H. Vucetich, Phys. Rev. {\bf D41}, 1034 (1990).
\bibitem{bbn}
L. Bergstrom, S. Iguri and H. Rubinstein, Phys. Rev. {\bf D60}, 045005 (1999).
\bibitem{webb}
J.K. Webb {\em et al.}, Phys. Rev. Lett. {\bf 82}, 884 (1999).
\bibitem{webbnew}
J.K. Webb {\em et al.}, seminar at IoA, Cambridge, May 2000.
\bibitem{steen}
S. Hannestad, Phys. Rev. {\bf D60}, 023515 (1999);

M. Kaplinghat, R.J. Scherrer and M.S. Turner, Phys. Rev. {\bf D60},
023516 (1999).
\bibitem{us}
P.P. Avelino, C.J.A.P. Martins and G. Rocha, Phys. Lett. {\bf B483}, 210 (2000).
\bibitem{boold}
P.D. Mauskopf {\it et al.}, Ap. J. 536, L59 (2000).
\bibitem{ourfit}
P. P. Avelino, C. J. A. P. Martins, G. Rocha and P. Viana, `Looking for a
Varying $\alpha$ in the Cosmic Microwave Background', Preprint DAMTP-2000-88,
submitted to Phys. Rev. D (2000).
\bibitem{bass}
B. A. Bassett, S. Liberati, C. Molina-Paris and M. Visser, astro-ph/0001441 (2000).
\bibitem{vsh}
A. Vilenkin \& E. P. S. Shellard, {\it Cosmic Strings and other Topological
Defects}, (Cambridge University Press: Cambridge, 1994).
\bibitem{linde}
A.D. Linde, {\it Particle Physics and Inflationary Cosmology} (Harwood, Chur,
Switzerland, 1990).
\bibitem{gun}
A. Albrecht {\em et al.}, Phys. Rev. Lett. {\bf 76}, 1413 (1996);

J. Magueijo {\em et al.}, Phys. Rev. Lett. {\bf 76}, 2617 (1996);

W. Hu and M. White, Phys. Rev. Lett. {\bf 77}, 1687 (1996).
\bibitem{turok}
N. Turok, Phys. Rev. {\bf D54}, 3686 (1996);

N. Turok, Phys. Rev. Lett. {\bf 77}, 4138 (1996).
\bibitem{hor}
A. Liddle, Phys. Rev. {\bf 51}, 5347 (1995).
\bibitem{also}
Y. Hu, M. S. Turner and E. J. Weinberg, Phys. Rev. {\bf D49}, 3830 (1994).
\bibitem{desi}
D. Salopek, J. R. Bond and J. M. Bardeen, Phys. Rev. {\bf D40}, 1753 (1989).
\bibitem{liddle}
A. Liddle, astro-ph/9910110 (1999);

J. Barrow and A. Liddle, Gen. Rel. Grav. {\bf 29}, 1501 (1997).
\bibitem{alex}
A. Lewin and A. Albrecht, astro-ph/9908061 (1999).
\bibitem{gauss1}
P. Ferreira, J. Magueijo and K. M. Gorski, Ap. J. {\bf 503}, L1 (1998);

J. Pando, D. Valls-Gabaud and L.-Z. Fang, Phys. Rev. Lett. {\bf 81}, 4568 (1998).
\bibitem{gauss2}
A. J. Banday, S. Zaroubi and K. M. Gorski, astro-ph/9908070 (1999).
\bibitem{ASWA2}
P. P. Avelino, E. P. S. Shellard, J. H. P. Wu and B. Allen, 
Ap. J. Lett. {\bf 507}, L101 (1998).
\bibitem{isoc}
A. D. Linde, Phys. Lett. {\bf B158}, 375 (1985);

L. Kofman, Phys. Lett. {\bf B173}, 400 (1986);

A. D. Linde and V. Mukhanov, Phys. Rev. {\bf D56}, 535 (1997).
\bibitem{acm2}
P. P. Avelino, R. R. Caldwell and C. J. A. P. Martins, Phys. Rev. {\bf D59}, 123509, (1999).
\bibitem{cb}
C. Contaldi, M. Hindmarsh and J. Magueijo, Phys. Rev. Lett. {\bf 82}, 2034 (1999);

R. A. Battye and J. Weller, phys. Rev. {\bf D61}, 043501 (2000).
\bibitem{ms1}
C.J.A.P. Martins and E.P.S. Shellard, Phys. Rev. {\bf D53}, 575 (1996).
\bibitem{ms2}
C.J.A.P. Martins and E.P.S. Shellard, Phys. Rev. {\bf D54}, 2535 (1996).
\bibitem{thesis}
C.J.A.P. Martins, {\em Quantitative String Evolution}, Ph.D. Thesis,
University of Cambridge (1997).
\bibitem{model}
C. J. A. P. Martins, E. P. S. Shellard and B. Allen, `Extending the Velocity-dependent One-scale String Evolution Model', hep-ph/0003298,
submitted to Phys. Rev. D (2000).
\bibitem{carter}
B. Carter, in {\em  Tlaxcala lecture notes, 2nd Mexican School on Gravitation
and Mathematical Physics}, A. Garcia {\em et al.} (eds.) (1996).
\bibitem{gauu}
P. P. Avelino and C. J. A. P. Martins, astro-ph/0006303 (2000).
\end{references}
\end{document}